\documentclass[usenatbib]{mn2e}
\usepackage{epsfig,natbib}
\usepackage{amssymb}
\usepackage{multirow}
\usepackage{multicol}
\usepackage[varg]{txfonts}
\usepackage{color}

\newcommand{\fref}[1]{Fig.~\ref{#1}}
\newcommand{\frefs}[2]{Figs~\ref{#1} and \ref{#2}}
\newcommand{\tref}[1]{Table~\ref{#1}}
\newcommand{\eref}[1]{equation~\ref{#1}}
\newcommand{\cref}[1]{Chapter~\ref{#1}}
\newcommand{\sref}[1]{section~\ref{#1}}

\def \tc{t_{\mathrm{cool}}}
\def \cs{c_{\mathrm{s}}}
\def \msolar{\; M_{\odot}}
\def \msolaryr{\; M_{\odot} \; \mathrm{yr}^{-1}}

\begin{document}
\title[Resolved images of circumstellar structure]{Resolved images of
  self-gravitating circumstellar discs with ALMA}

\author[P.~Cossins, G.~Lodato \& L.~Testi] 
{
\parbox{5in}{Peter Cossins$^1$, Giuseppe Lodato$^2$ and Leonardo Testi$^3$}
\vspace{0.1in} 
 \\ $^1$ Department of Physics \& Astronomy, University of Leicester,
 Leicester LE1 7RH UK  
 \\ $^2$ Dipartimento di Fisica, Universit\`{a} Degli Studi di Milano, Via 
 Celoria 16, 20133, Milano, Italy
 \\ $^3$ European Southern Observatory, Karl Schwarzschild str. 2, D-85748
 Garching, Germany 
}

\maketitle

\begin{abstract}
  In this paper we present simulated observations of massive self-gravitating
  circumstellar discs using the Atacama Large Millimetre/sub-millimetre Array
  (ALMA).  Using a smoothed particle hydrodynamics model of a $0.2\msolar$
  disc orbiting a $1\msolar$ protostar, with a cooling model appropriate for
  discs at temperatures below $\sim 160$K and representative dust opacities,
  we   have constructed maps of the expected emission at sub-mm
  wavelengths. We have then used the CASA ALMA simulator to generate
  simulated images and visibilities with various array configurations and
  observation frequencies, taking into account the expected thermal noise
  and atmospheric opacities.  We find that at 345 GHz (870 $\mu$m) spiral
  structures at a resolution of a few AU should be readily detectable in
  approximately face-on discs out to distances of the Taurus-Auriga
  star-forming complex.  
\end{abstract}


\footnotetext[1]{E-mail: peter.cossins@astro.le.ac.uk}

\section{Introduction}
\label{intro}
Circumstellar discs play an important role in the formation and evolution of
both stars and planets, and as such have been the object of much study and
observation in recent years.  Millimetre wavelength surveys of star forming
regions in Taurus \citep{Beckwithetal90,Kitamuraetal02,AndrewsWilliams05},
Orion \citep{Eisneretal08} and $\rho$ Ophiucus
\citep{AndreMontmerle94,AndrewsWilliams07a} have provided extensive evidence of
circumstellar discs, while optical images have been provided by the Hubble
Space Telescope (HST) of HH30 \citep{Burrowsetal96} and the Fomalhaut system
\citep{Kalasetal08}.  However, relatively few systems have been imaged with
great enough resolution to determine the disc structure on scales of less than
a few tens of $AU$. 

This may be about to change however, with the Atacama Large
Millimetre/sub-millimetre Array (ALMA) due to come on line in the near future.
With a minimum beam diameter of approximately 5 milli-arcseconds at 900
GHz, ALMA should ideally provide resolution down to $\approx 2 AU$ for discs
observed in Orion, with sub-$AU$ resolution for systems in Taurus-Auriga.  It
will therefore allow observations of disc sub-structure in unprecedented
detail.  The discovery of structures within discs could have important
implications for our understanding of the evolution of both the discs
themselves \citep{RiceA09} and the protostars they orbit.  

Gaseous circumstellar discs are expected to be turbulent
\citep{Ebert94,Gammie96}, with the internal stresses that this turbulence
induces being the principal driver of angular momentum transport, and thus
accretion.  This turbulence will lead to sub-structure being
present at all scales within the disc, with those on smaller scales (due to
the magneto-rotational instability) remaining undetectable, while the larger
scale structure induced by the gravitational instability should be resolvable
with ALMA.   

It is expected that protostellar discs about Class 0/Class I objects will go
through a self-gravitating phase
\citep{LodatoBertin01a,VorobyovBasu05c,Hartmannbook} as the infall rate from
the gaseous envelope is much greater than the accretion rate onto the
protostar \citep{VorobyovBasu06}.  In this phase large amplitude
spiral structures will form, driving accretion onto the protostar
\citep{LinPringle87,LaughlinBodenheimer94,ArmitageLP01,LodatoR04,LodatoR05}
and potentially leading to the formation of companion objects.  These
companions may vary in size from stellar mass
\citep{Bonnell94,BonnellBate94,Whitworthetal95} to brown dwarf (BD) or planetary mass \citep{RiceLA05,StamatellosHW07,Clarke09,Cossinsetal10}.  It is
possible that protoplanetary discs around Classic T-Tauri (Class II) stars may
also exhibit spiral patterns due to the presence of gravitational
instabilities
\citep{LodatoBertin01a,Boleyetal06,VorobyovBasu08,Cossinsetal10}, and indeed
there are already possible detections of spiral structures in the discs of GSS
39 in Ophiuchus \citep{Andrewsetal09} and in IRAS 16293-2422B
\citep{Rodriguezetal05}.  

At radii greater than approximately $50 AU$, planet formation through
direct fragmentation of these spiral overdensities into bound objects is
possible \citep{Boss97,Boss98,Clarke09,Rafikov09,Cossinsetal10}.  While
\citet{WolfD'Angelo05} have indicated that the forming giant (proto-)planets themselves may be
detectable using ALMA, the observability of the large-scale spiral structure
itself within protostellar and protoplanetary discs, though implied
\citep{TestiS08}, has remained undemonstrated.  In this paper we therefore
present a simple self-gravitating disc simulation, and from it we derive mock
observations of disc systems at the resolutions and sensitivities that should
be possible with ALMA.  Hence in \sref{simulations} we detail the simulation,
and in \sref{generation} we discuss how the mock observations are generated
from it, taking into account telescope effects and sensitivities.  In
\sref{results} we present the observations for various system and telescope
parameters, and finally in \sref{discussion} we discuss the significance of
our results.    
\section{Numerical Simulation of Structure Formation}
\label{simulations}

Gravitational instabilities in discs have been extensively studied numerically
using both grid-based Eulerian codes
\citep{Gammie01,Pickettetal03,Boleyetal06,Kratteretal10} and Lagrangian
particle-based methods
\citep{Riceetal03,LodatoR04,LodatoR05,NayakshinCS07,Alexanderetal08,ForganRice09,Cossinsetal09}
and the reader is referred to these for further details of the dynamics of
gravitationally unstable discs.  The simulation we use to generate the mock
observations was performed in the manner detailed in \citet{Cossinsetal09}
using a 3D smoothed particle hydrodynamics (\textsc{sph}) code, a Lagrangian
hydrodynamics code capable of modelling self-gravity \citep[see for
example,][]{Benz90,Monaghan92,Monaghan05,PriceMonaghan07}.  Our system
consists of a single 
point mass orbited by 500,000 \textsc{sph} gas particles, with the central
object free to move under the gravitational influence of the disc.  All
particles evolve according to individual time-steps governed by the Courant
condition, a force condition \citep{Monaghan92} and an integrator limit
\citep{BateBP95}.  

We allow the discs to cool towards gravitational instability by implementing a
simple cooling law of the form 
\begin{equation} 
  \frac{du_{{i}}}{dt} = - \frac{u_{i}}{t_{\mathrm{cool},i}},
  \label{coolinglaw}
\end{equation}
where $u_{i}$ and $t_{\mathrm{cool},i}$ are the specific internal energy and
cooling time associated with the $i^{th}$  particle.  We then determine the
cooling time $\tc$ through the simple prescription that 
\begin{equation}
  \Omega \tc = \beta
  \label{cooling}
\end{equation}
where $\Omega$ is the angular frequency and $\beta$ is a fixed parameter
throughout each simulation.  Although this is clearly an \emph{ad hoc} cooling
function, 
it can be used as a simple parameterisation in order to conduct controlled
numerical experiments.  In \citet{Cossinsetal09} we found that for a given
disc mass, the spiral structures formed through the gravitational
instability (as characterised by the radial and azimuthal wavenumbers of the
excited modes) are \emph{independent} of the cooling, but that the strength of
the modes (characterised by the relative RMS amplitude of the surface density
perturbations $ \left< \delta \Sigma / \Sigma \right>$) varies such that 
\begin{equation}
  \left< \frac{\delta \Sigma}{\Sigma} \right> \approx
  \frac{1}{\sqrt{\Omega \tc}}.
\end{equation}
Therefore, notwithstanding the simplicity of our cooling prescription, we may
reasonably assume that the spiral structures formed in physical systems with
characteristic masses corresponding to those in our simulation will have
qualitatively similar structure formation, with the uncertainty lying
primarily in the amplitudes of the density perturbations. 

Furthermore, in \citet{Cossinsetal10} we characterise the behaviour of
$\Omega \tc$ with radius, and show that for accretion rates $\sim 10^{-7}
\msolaryr$ (typical for Classic T Tauri objects) at radii of $20 - 50 AU$ the
value of $\Omega \tc$ is $\sim 10^{1} - 10^{2}$, decreasing with increasing
radius.  Hence although the cooling formalism given in \eref{cooling} is very
simple, it produces spiral structure in the correct modes and at approximately
the correct amplitudes for the outer radii of discs, as long as we choose
$\beta$ in the range $\gtrsim 10$.  It is therefore useful as a means of
generating test cases to investigate whether such structures would actually be
observable, particularly in the outer parts of discs.

\subsection{Simulation Details}
The simulation used to generate the mock observations described here consists
of a $0.2\msolar$ disc about a $1.0\msolar$ mass star, extending out to
approximately 25 AU.  Although this is a relatively high mass ratio, it is
within observed bounds \citep{AndrewsWilliams05,AndrewsWilliams07a}, and is 
plausible for the early (Class I) stages of intermediate mass protostellar
evolution, when the disc is expected to be self-gravitating
\citep{LodatoBertin01a,VorobyovBasu05c,Hartmannbook}. 

Our initial conditions consist of a disc of gas particles on circular
orbits, distributed with a surface density profile $\Sigma \sim R^{-\gamma}$
with $\gamma = 1.5$, as for the Minimum Mass Solar Nebula (MMSN,
\citealt{Weidenschilling77}).  Note that observational constraints for
$\gamma$ range from $0.4 \lesssim \gamma \lesssim 1.0$ for protoplanetary
discs in Ophiuchus \citep{Andrewsetal09} to $0.1 \lesssim \gamma \lesssim 1.7$
in Taurus \citep{AndrewsWilliams07b}, so this value is within observed limits.

The disc is initially in approximate vertical hydrostatic equilibrium with a
Gaussian distribution of particles where the scale height $H = \cs /
\Omega$ and $\cs$ is the sound speed. The azimuthal velocities take into
account both a pressure correction \citep{LodatoRNC07} and the enclosed disc
mass and any variation from dynamical equilibrium is washed out on the
dynamical timescale.  The initial temperature profile is such that $\cs^{2}
\sim R^{-1/2}$, with the minimum value of the Toomre parameter
$Q_{\mathrm{min}} = 2$ occurring at the outer edge of the disc. In this manner
the disc is initially gravitationally stable throughout its radial range.   

Note that the \textsc{sph} code and the initial conditions used are exactly
the same as were used in \citet{LodatoR04,LodatoR05} and
\citet{Cossinsetal09,Cossinsetal10}, excepting the fact that here we have
used a mass ratio of 0.2, and the reader is directed to these for further
details.  Finally, in terms of cooling, we set $\beta = 7$ and use the cooling
formalism described in \eref{coolinglaw} above.  This is approximately in the
expected range, and is low enough to avoid spurious numerical heating
effects due to artificial viscosity \citep{LodatoR04}.

\subsection{Disc Evolution}
\begin{figure}
  \centering
  \includegraphics[angle=270,width=20pc]{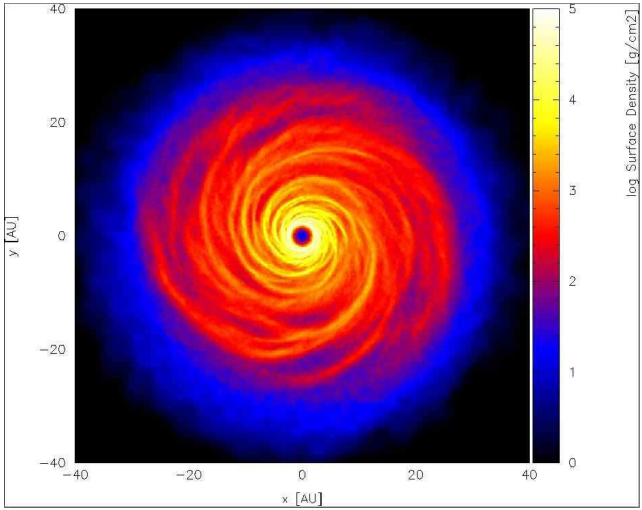}
  \caption{Simulated surface density perturbations in a 0.2 $\msolar$ disc
    about a 1.0 $\msolar$ protostar.  The gravitationally induced spiral waves
    that impart heat to the disc are clearly visible.} 
  \label{Surfacedensity}
\end{figure}

Although the thermal profile of the disc initially ensures it is gravitationally
stable at all radii, it is however \emph{not} in thermal equilibrium.  As the
simulation evolves, the disc therefore cools towards gravitational
instability, which is initiated when $Q \approx 1$, after approximately 1000
years.  The disc then settles into a marginally
stable, quasi-steady dynamic thermal equilibrium state, characterised by the
presence of spiral density waves that propagate across the face of the disc,
and which provide heat (through shocks) to balance the imposed cooling.  These
spiral waves are clearly seen in the surface density, as shown in
\fref{Surfacedensity}.  The simulation was run for approximately 10,000 years
(equivalent to $\sim 10$ cooling times at the disc outer edge), suggesting
that during the self-gravitating period of protostellar evolution (expected to
last a few $ \times 10^{5}$ years during and immediately after the
infall phase) circumstellar discs should indeed be able to reach this
quasi-steady state. Since the surface density varies on the longer
viscous timescale, the original $\Sigma \sim R^{-3/2}$ profile remains
largely unchanged throughout the simulation. 

\begin{figure}
  \centering
  \includegraphics[width=20pc]{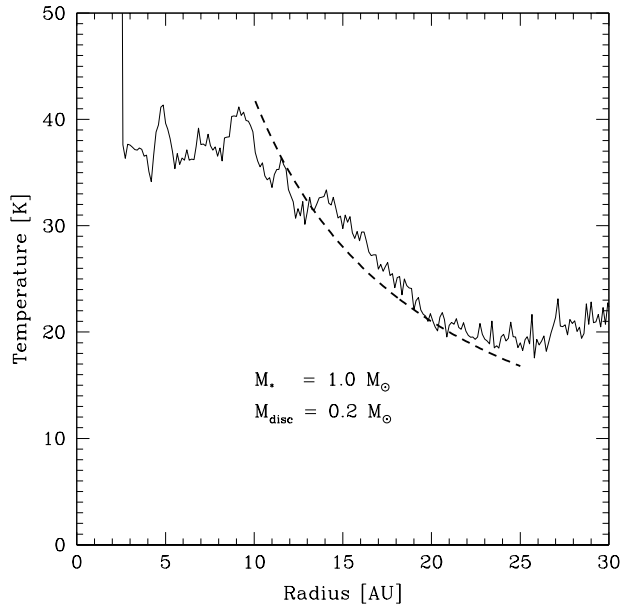}
  \caption{Azimuthally averaged temperature profile of the disc once it has
    settled into the dynamic thermal equilibrium state.  The dashed
    line shows an $R^{-1}$ profile.}
  \label{temperature}
\end{figure}

Once the disc reaches the self-regulated dynamic thermal equilibrium state,
the disc temperature falls to approximately 20 - 40 K, with the minimum at the
disc outer edge, as shown in \fref{temperature}.  Between roughly 10
and 25 AU the temperature falls off as $R^{-1}$, in reasonable
agreement with observations \citep{AndrewsWilliams07b}.  Note that the slight
temperature \emph{rise} outside of $R \approx 25$ AU is due to very
low density, higher temperature gas external to the main bulk of the disc.

\section{Generation of mock observations}
\label{generation}
Having run the hydrodynamic simulations, it is then necessary to use the disc
parameters to create flux maps of the objects as they may appear using ALMA.
Throughout the following we assume that the disc is face-on to the observer.

In order to calculate the emission we use the individual particle densities
$\rho_{\mathrm{i}}$ generated from the \textsc{sph} simulations, and calculate
the particles' absorption coefficients $\alpha_{\nu,\mathrm{i}}$ at frequency
$\nu$ using 
\begin{equation}
  \alpha_{\nu,\mathrm{i}} = \rho_{\mathrm{i}} \kappa_{\nu},
  \label{alphanu}
\end{equation}
where $\kappa_{\nu}$ is the opacity of the disc at frequency $\nu$.

For a face-on disc the optical depth $\tau_{\nu}$ at frequency $\nu$ is
defined in the following manner 
\begin{equation}
  \tau_{\nu} = \int_{-\infty}^{\infty} \alpha_{\nu}(z) \; \mathrm{d}z,
\end{equation} 
and hence for a disc of SPH particles, we may evaluate the vertical optical
depth at any point on the disc face by approximating this integral as a sum
over all the relevant interacting particles.  We therefore use the following
approximation for the optical depth of the disc as seen by a distant observer
\begin{equation}
  \tau_{\nu} \approx \sum_{\mathrm{i}} \rho_{\mathrm{i}} \kappa_{\nu}
  w_{\mathrm{i}} 
  \label{tau}
\end{equation}
where $w_{\mathrm{i}}$ is a weighting function related to the particle's mass,
density, and the \textsc{sph} smoothing kernel (see \citealt{PriceSplash}) and
where $i$ denotes all particles along a given line of sight through the disc.

We assume that radiation from the disc is in thermal equilibrium with itself
and thus that the disc emits as a black body, with the source function
$S_{\nu}$ at frequency $\nu$ given by the Planck function,
\begin{equation}
  S_{\nu} = B_{\nu}(T) = \frac{2 h \nu^{3}}{c^{2}} \frac{1}{e^{h\nu / kT} - 1},
  \label{Planck}
\end{equation}
where $h$ is Planck's constant, $c$ is the speed of light \textit{in vacuo},
$k$ is Boltzmann's constant and $T$ is the temperature of the emitter.
For simplicity, we assume that our discs are vertically isothermal with
temperature $T$, (which we obtain via a vertical average from the
simulations) meaning that at each point in the disc the source function is
constant with height, and is given by $B_{\nu}(T)$.  Modelling the
disc in this manner, as a geometrically thin structure with no
vertical temperature gradient, is obviously a very crude
approximation. Nevertheless, since we are only interested in
predicting the emission in the (sub-)millimetre range, this
approximation is justified as the bulk of the emission will indeed
come from the relatively isothermal layer located about the disc 
midplane. 

For such a constant
source function, the specific intensity (or surface brightness) $I_{\nu}$ at
frequency $\nu$ and optical depth $\tau_{\nu}$ is given by   
\begin{equation}
  I_{\nu} = B_{\nu}(T) \left( 1 - e^{-\tau_{\nu}} \right).
  \label{brightness}
\end{equation}
Note that this formulation for the surface brightness is often used in
reverse to infer disc properties from the (sub) millimetre surface
brightness, e.g. \citet{Beckwithetal90,Andrewsetal09}.  Furthermore, it is
also clear that once the disc becomes optically thick its emission is
determined solely by its temperature, whereas in optically thin regions the
emission is exponentially dependent on $\tau_{\nu}$.  Optically thick
structures forming in optically thin parts of the disc are therefore
likely to show greater variation in intensity (and thus be more
readily observable) than structure in purely optically thick regions. 

\begin{figure*} 
  \centerline{
    \includegraphics[angle=270,width=13pc]{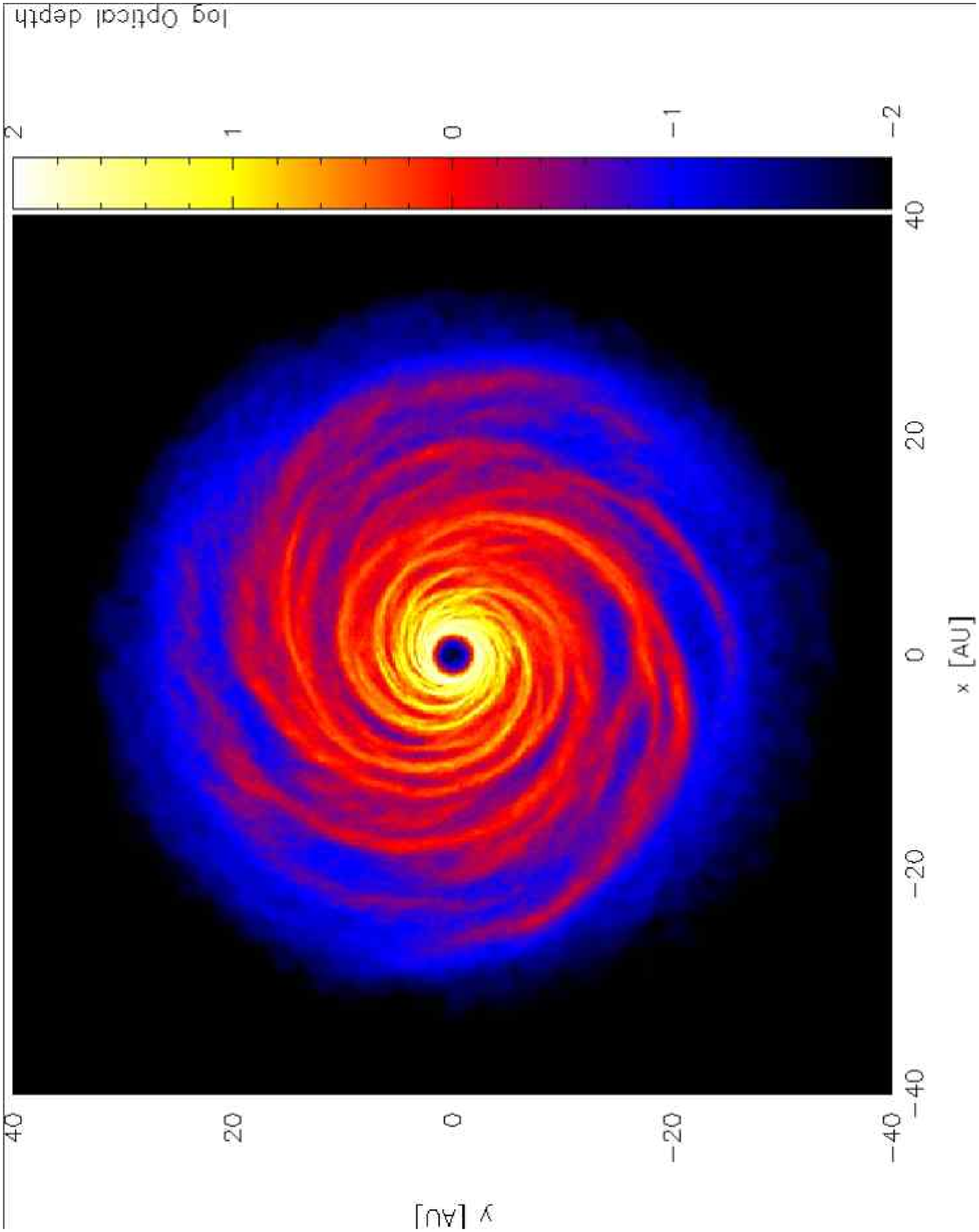}
    \includegraphics[angle=270,width=13pc]{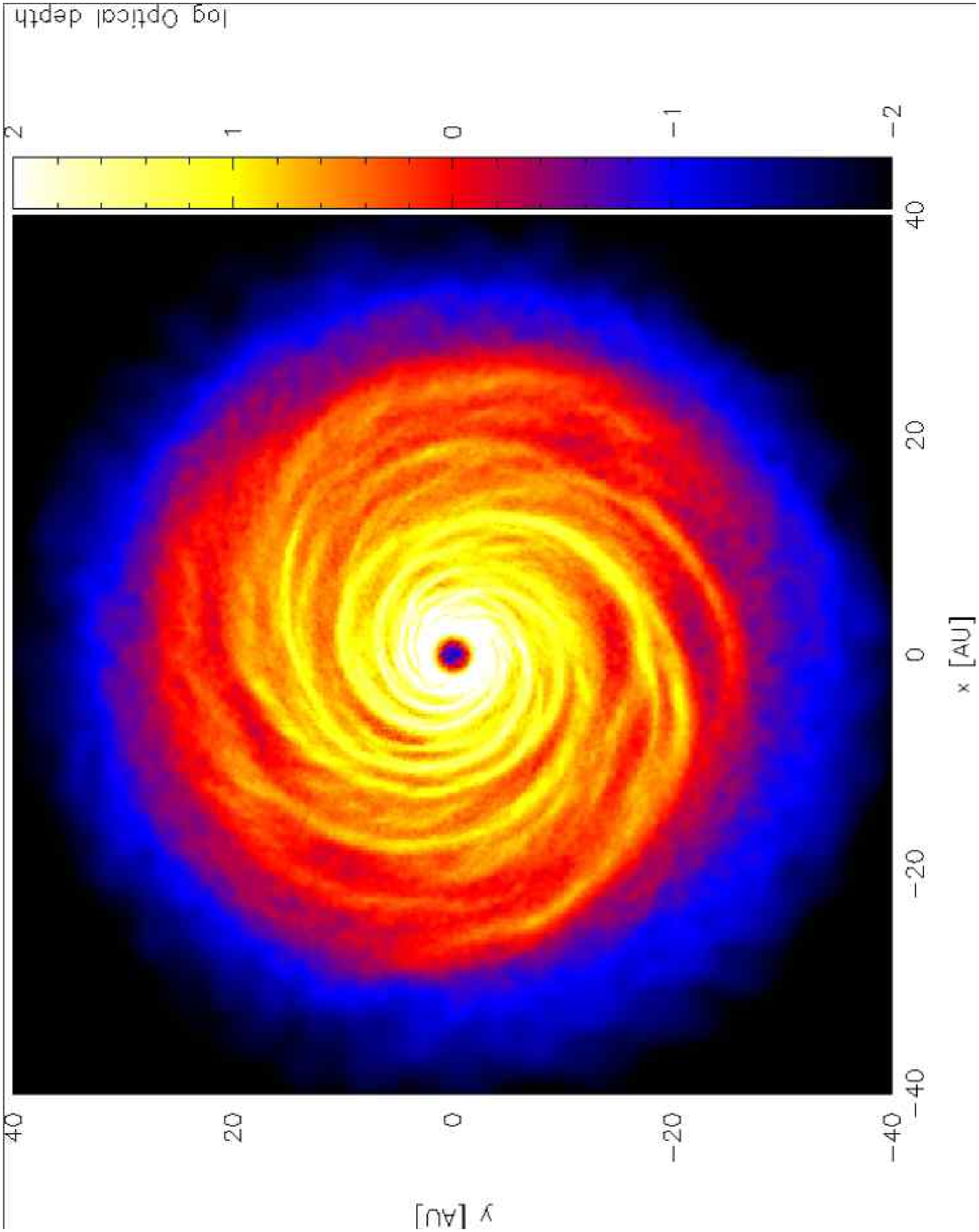}
    \includegraphics[angle=270,width=13pc]{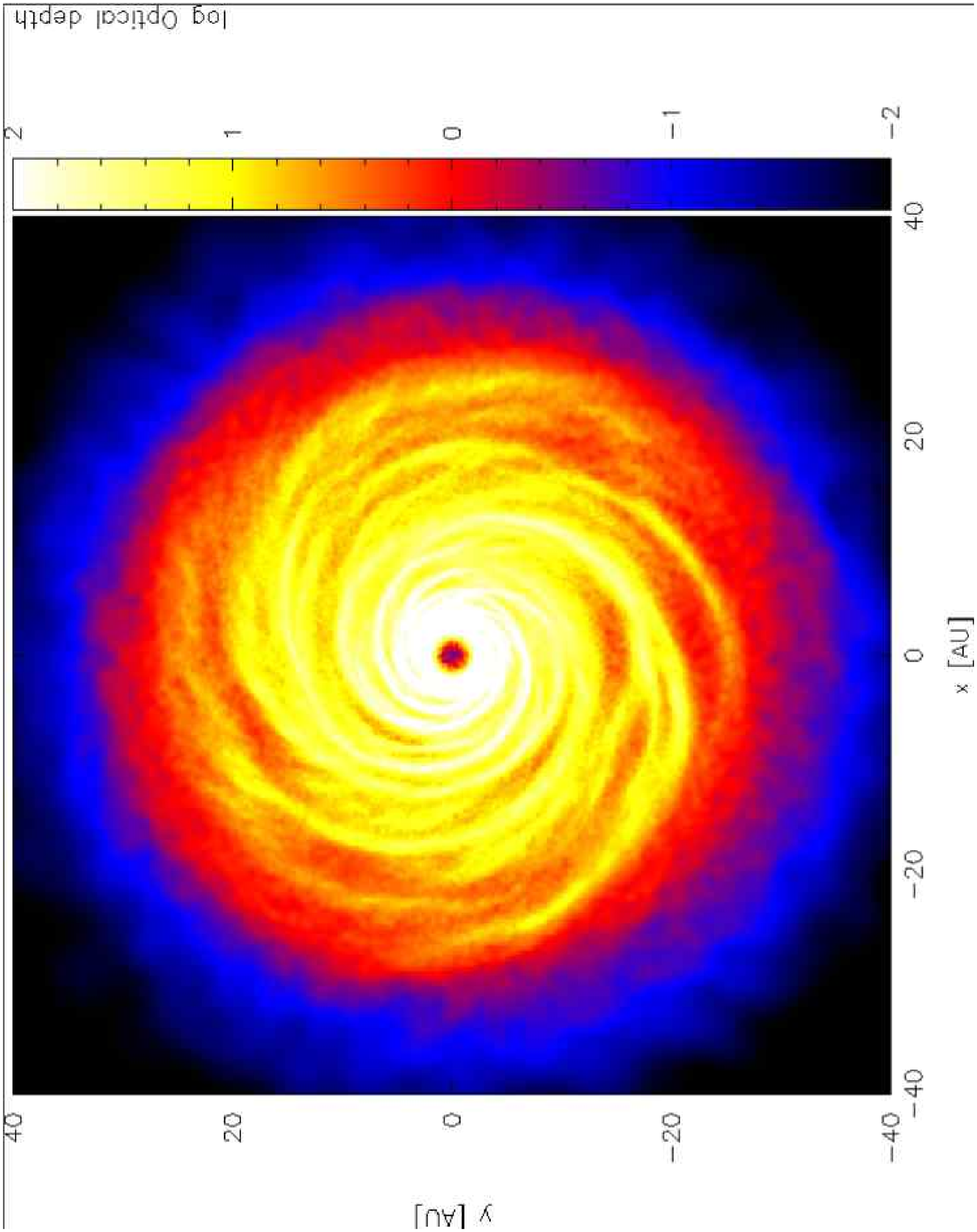}
  }
  \caption{Comparison of the (logarithmic) optical depth across the
    disc face at 
    frequencies of 45 GHz (left), 345 GHz (centre) and 870 GHz (right).  The
    greatest contrast between optically thick and optically thin regions
    across the arm/inter-arm regions in the bulk of the disc is to be seen in
    the 345 GHz case.} 
  \label{opticaldepth}
\end{figure*}  

\subsection{Dust opacities}
\label{opacities}
A critical parameter in the above calculation is the opacity of the disc
$\kappa_{\nu}$.  At temperatures below $\approx 160 K$ the Rosseland mean
opacity becomes dominated by ices \citep{BellLin94,Cossinsetal10}, but
the specific value of the opacity at a given frequency $\kappa_{\nu}$ is
determined by the grain size distribution \citep{MiyakeNakagawa93}.  

To determine the value of $\kappa_{\nu}$ we use the power-law model of
\citet{Beckwithetal90}, such that 
\begin{equation}
  \kappa_{\nu} = \kappa_{12} \left( \frac{\nu}{\nu_{12}} \right)^{\beta},
\end{equation}
where $\nu_{12} = 10^{12}$ Hz, and where $\kappa_{12}$ is the value of
the opacity at this fiducial frequency.  The normalisation constant
$\kappa_{12}$ is relatively poorly constrained, with values ranging from 0.1
cm$^{2}$ g$^{-1}$ \citep{Beckwithetal90} to approximately 0.016
\citep{Krameretal98,Rafikov09}.  For protoplanetary and protostellar discs the
power-law index $\beta$ can be determined from the disc SED in the mm/sub-mm
range, and is found to be in the region $0.3 \lesssim \beta \lesssim 1.5$
\citep{Kitamuraetal02,Testietal03,Riccietal09}. Note that this range is below
that expected for the interstellar medium $\beta \approx 1.7 - 2$
\citep{Finkbeineretal99,ChakrabartiMcKee08,Hartmannbook}, which is
attributed to grain processing within the disc.  

For our models we use a fiducial dust opacity $\kappa_{12} = 0.025$ cm$^{2}$
g$^{-1}$, and in accordance with \citet{Rafikov09} we assume throughout that
$\beta = 1.0$, such that 
\begin{equation}
    \kappa_{\nu} = 0.025 \left(\frac{\nu}{10^{12} \; \mathrm{Hz}} \right) \;
    \mathrm{cm}^{2} \; \mathrm{g}^{-1}. 
  \label{opacity}
\end{equation}

\subsection{The ALMA Simulator}
\label{salmasim}

To produce realistic simulated ALMA observations of the models we have
used the ALMA simulator {\tt simdata} in CASA\footnote{CASA (Common Astronomy
Software Applications) is being primarily developed by ALMA and the
National Radio Astronomy Observatory (NRAO) as the main 
off-line data reduction package for ALMA and EVLA, the Expanded Very
Large Array. {\tt http://casa.nrao.edu}}.
The version of the simulator we used (2.4) allows us to use the latest version
of the planned ALMA antenna configurations, the expected receiver noise based
on technical specifications and the contribution due to the
atmosphere, itself
based on input values for the atmospheric temperature
$T_{\mathrm{atm}}$ and the optical depth at the frequency of the
simulations.  In all the simulations presented here we assume
$T_{\mathrm{atm}}=265$~K, and the optical depth is computed
using the ATM atmosphere models of \citet{Pardoetal02}.  We
use typical Chajnantor conditions and an amount of precipitable water
vapour (pwv) as expected for dynamical scheduling of the observations.
In \tref{tsimul} we therefore show the water content and
opacity of the atmosphere, the expected theoretical noise and the
angular and linear resolution of the 
simulations for discs at 50~pc and 140~pc for each
frequency.  

Note that we have also run simulations for the lowest (45~GHz) and highest
(870~GHz) planned ALMA bands -- contrary to the intermediate
frequencies, these frequency bands will not be available when ALMA
first comes on-line. High frequencies receivers are
planned to be introduced during the early years of ALMA science
operations, while the low frequency band is under discussion for the ALMA
development programme. 
 
The array configuration used varied for each frequency, as we
aimed to use the configuration that offered the best compromise
between surface brightness sensitivity and angular resolution.  Note
that at low frequencies the angular resolution is limited by the largest
available array configuration.  In all cases we performed aperture
synthesis simulations with transit duration of 2~hrs. 

\begin{table*}
  \begin{center}
    \begin{tabular}{crlcrlrl}
      \hline
      \textbf{Frequency} & \multicolumn{2}{c}{\textbf{Atmospheric Conditions}} & 
      \textbf{Expected Sensitivity} &
      \multicolumn{2}{c}{\textbf{Resolution at 50pc}} &
      \multicolumn{2}{c}{\textbf{Resolution at 140pc}} \\
      (GHz) & pwv (mm) & $\tau_0$ & 1$\sigma$ ($\mu$Jy/beam) &
      (mas) & (AU) & (mas) & (AU) \\
      \hline
      45  & 2.3 & 0.05 & 3   & 120 & 6   & 120 & 16.8 \\
      100 & 2.3 & 0.03 & 4   & 50  & 2.5 & 50  & 7    \\
      220 & 2.3 & 0.1  & 10  & 60  & 3   & 20  & 2.8  \\
      345 & 1.2 & 0.2  & 20  & 40  & 2   & 25  & 3.5  \\
      680 & 0.5 & 0.6  & 60  & 50  & 2.5 & 35  & 4.9  \\
      870 & 0.5 & 0.7  & 100 & 60  & 3   & 45  & 6.3  \\
      \hline
    \end{tabular}
  \end{center}
  \caption{Atmospheric conditions, expected sensitivities and angular
    and linear resolutions as a function of frequency for the simulated
    observations, assuming distances of 50 pc (corresponding to the
    distance of the TW Hya association) and 140 pc (roughly
    corresponding to the Taurus-Auriga, Ophiucus and Chamaeleon star
    forming regions). The amount of precipitable water vapour (pwv, in
    mm) has been chosen according to the current ALMA dynamical
    scheduling expectations. Note that at the higher frequencies the
    varying resolutions at different distances is due to the fact that
    we have considered different ALMA configurations in order to
    obtain an optimal compromise between resolution and
    sensitivity. On the other hand, at low frequency, the resolution
    is dictated by the largest available array configuration, and thus
    remains constant with distance.}  
  \label{tsimul} 
\end{table*}

\section{Results}
\label{results}

Using the disc simulation shown in \fref{Surfacedensity}, we investigated the
emission at the various frequencies shown in \tref{tsimul}.  Based on the
requirement that the optical depth should vary between optically thick in the
spiral arms and optically thin in the inter-arm regions (to maximise the
contrast in emission across these regions) we have principally considered the
results at 345 GHz.  Furthermore, at this frequency we expect the resolution
to be roughly 1 AU at both TW Hydrae and Taurus-Auriga distances, while the
telescope sensitivity should not be a limiting factor. To illustrate the
effects of varying the frequency, the optical depth at 45, 345 and 870 GHz is
shown in \fref{opticaldepth} for comparison.  Also clear from this
figure is the expected increase in optical depth with frequency due to
increasing opacity. 

\subsection{Simulated ALMA Images}

\begin{figure*}
  \centering
  \includegraphics[width=20pc,angle=270]{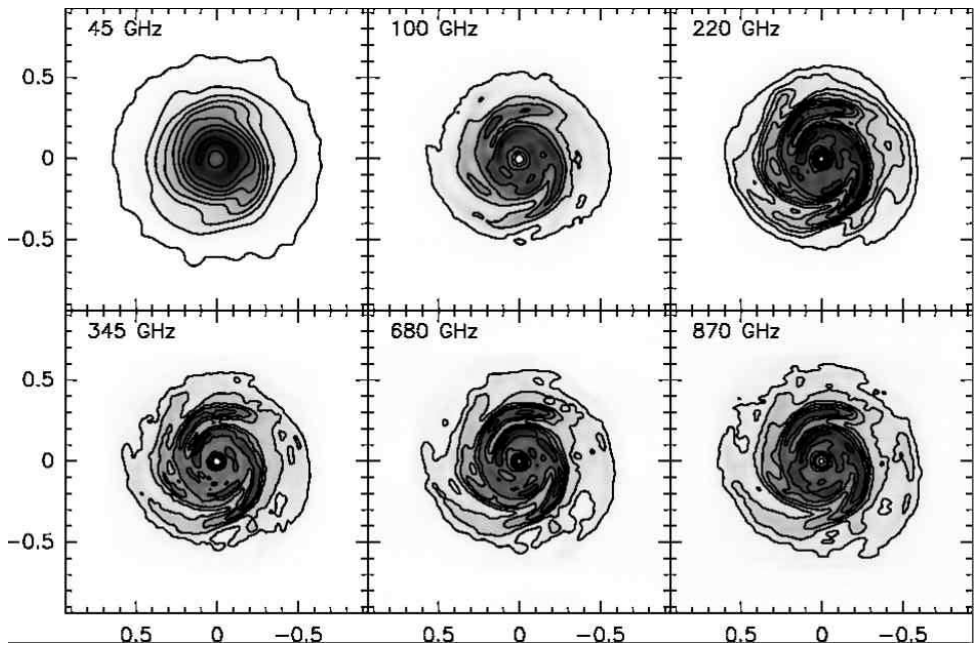}
 \caption{Simulated aperture synthesis ALMA images for a disc at
    50~pc, with a transit duration of 2 hours. From left to right and
    top we show simulations computed for an observing frequency of 45,
    100, 220 (top) and 345, 670 and 870~GHz (bottom). Axis scales are
    in arcseconds. Contours start at 0.01 and are spaced by
    0.08~mJy/beam at 45~GHz, start at 0.08 and are spaced by
    0.2~mJy/beam at 100~GHz, start at 0.5 and are spaced by
    0.5~mJy/beam at 220~GHz, start at 0.8 and are spaced by
    0.8~mJy/beam at 345~GHz, start at 2 and are spaced by 2~mJy/beam
    at 680~GHz, start at 4 and are spaced by 4~mJy/beam at 870~GHz.}
  \label{fsimul50}
\end{figure*}

Having generated the specific intensity map by the method described in
\sref{generation}, we have used them to 
simulate ALMA observations as described in \sref{salmasim}.
In \frefs{fsimul50}{fsimul140} we therefore show the results of our
simulations for discs at a distance of 50pc (TW~Hya) and at 140pc (Chamaeleon,
Ophiuchus, Taurus-Auriga). Note that our simulations only include
the effects of thermal noise from receivers and the atmosphere, but do
not take into account calibration uncertainties and residual phase
noise after calibration. These effects are likely to be most important
at high frequencies and long baselines, so the simulated maps at 680
and 870~GHz, especially for the 140pc case, represent observations
carried out in ideal conditions and with excellent calibrations.

The simulations show that the predicted spiral structure is readily
detectable at all simulated frequencies at the 50 pc distance of the
TW~Hya association.  In the case of star-forming regions at 140pc the
situation is less clear cut.  At low frequencies 
($\lesssim 100$~GHz) even ALMA will probably not provide the angular
resolution required to image the spiral structure clearly, whereas at
the highest frequencies, as noted above, the simulations are probably
over optimistic.  Nevertheless, our simulations show that at 220 and
345~GHz (ALMA Bands 6 and 7), the predicted structures should be
clearly detectable. 

\begin{figure}
  \centering
  \includegraphics[width=15pc]{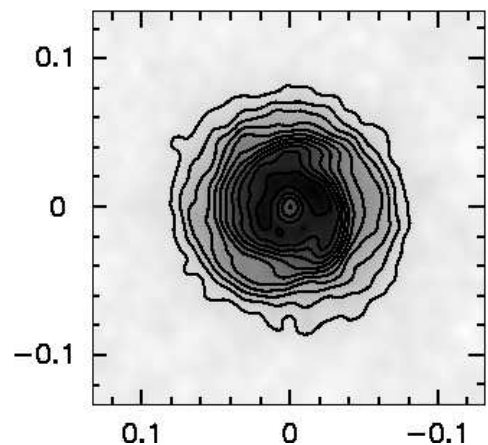}
  \caption{Simulated 345 GHz aperture synthesis image of a disc at 410~pc,
    with a transit duration of 2 hours.  Contours start at and are
    spaced by 60 mJy/beam.  Clear asymmetries are present, but the
    underlying spiral structure is not well resolved.}
  \label{fsimul410}
\end{figure}

Finally, in \fref{fsimul410} we show the predicted observability of a
disc at 410pc (equivalent to the Orion Nebula Cluster distance) imaged
at 345 GHz.  While the non-axisymmetric nature of the disc is clear,
the spiral structure \emph{per se} is not well resolved, and thus we
infer that even with ALMA such structures will not be conclusively
detectable. 

\begin{figure*}
  \centering
 \includegraphics[width=20pc,angle=270]{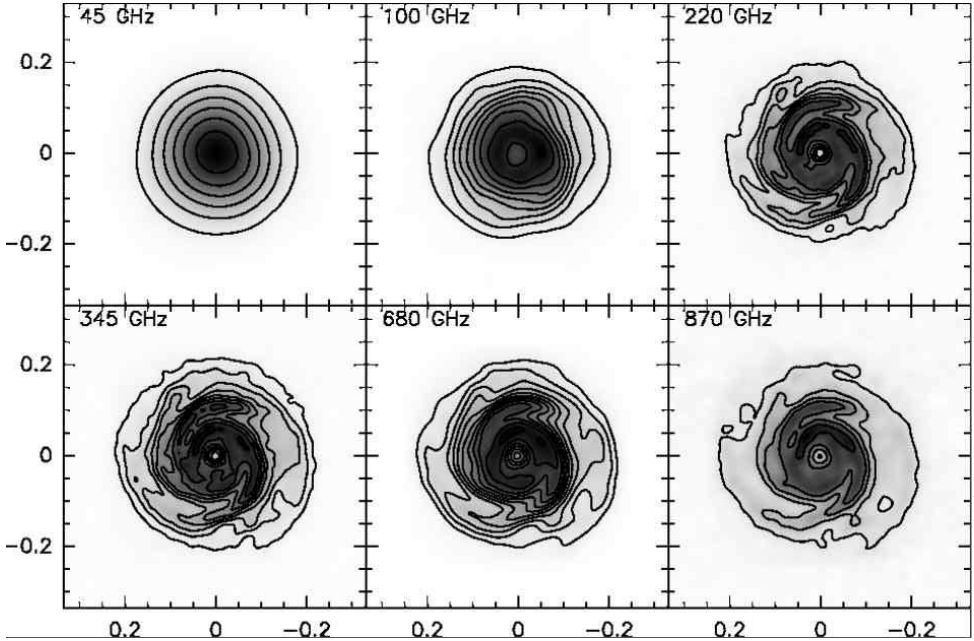}
  \caption{As for \fref{fsimul50}, but for a disc at 140~pc. Contours
    start at 0.08 and are spaced by 0.08~mJy/beam at 45~GHz, start at
    0.08 and are spaced by 0.08~mJy/beam at 100~GHz, start at 0.1 and
    are spaced by 0.1~mJy/beam at 220~GHz, start at 0.2 and are spaced
    by 0.2~mJy/beam at 345~GHz, start at 2 and are spaced by 1~mJy/beam
    at 680~GHz, start at 1 and are spaced by 1~mJy/beam at 870~GHz.}
  \label{fsimul140}
\end{figure*}

\section{Discussion}
\label{discussion}

In this paper, we have used a 3D, global SPH simulation of a massive
($0.2\msolar$) compact ($R_{\mathrm{out}} = 25$ AU) self-gravitating disc
about a young star ($1.0\msolar$) to demonstrate that the spiral modes excited
by the gravitational instability should be detectable in face-on circumstellar
discs using ALMA.  At distances comparable to the TW Hydrae association ($\sim
50$ pc), such spiral density waves are readily apparent with observation times
of 2 hours, whereas at Taurus distances of $\sim 140$ pc a careful
choice of the observing frequency and excellent observing conditions  may be
required for significant detections.  Our results suggest that structure in
such discs in Orion ($\sim 400$pc) will most likely not be resolvable.

In order to generate these predicted observations we have used temperature
and density maps provided by numerical simulations, together with an
empirical relationship for the dust opacity of circumstellar material to
obtain the optical depth at each part of the disc face.  In a precisely
similar manner to that used to obtain disc properties from sub-mm observations
\citep{Beckwithetal90} we have then been able to determine the specific
intensity of the disc emission across the disc face, which has then
been used as input for the ALMA simulator in CASA to produce simulated
observations.  We find that observation frequencies of 
around 345 GHz (870 $\mu$m) are ideal for this kind of observation, as we find
the spiral arms to be optically thick, whereas the inter-arm regions are
optically thin, maximising the contrast in emission between the regions.  It
should be noted however that this `ideal' frequency range is dependent on our
assumptions concerning the dust opacity, in which there is considerable
scatter. 

There are certain limitations to our model which should be noted however.  Our
stellar and disc masses are both at the upper end of the expected
distributions \citep{AndrewsWilliams05,AndrewsWilliams07a,Beckwithetal90},
although in the early stages of star formation (Class I objects) such high
disc masses are not unreasonable due to the infalling envelope.  Of necessity,
in order to be self-gravitating our discs are cold ($20 - 40$ K), which implies
both a relatively low ambient temperature (not unreasonable, since giant
molecular clouds tend to have temperatures of $\sim 10K$,
\citealt{Myersetal83,MyersBenson83}) and for heating from the protostar to be
negligible.  While this latter assumption is clearly unlikely to be valid for
the surface layers of the disc that are irradiated directly by the star, the
disc midplane (which dominates the emission) is likely to be cold enough to
justify this assumption
\citep{AndrewsWilliams05,Dullemondprpl,D'Alessioetal98}.  In a similar manner,
this assumption of a colder inner layer allows us to ignore the effects of the
magneto-rotational instability \citep{Gammie96}.  

Given that our disc is quite compact, and many discs are observed to
extend out to much larger radii ($\sim 10^{2} - 10^{3}$ AU,
\citealt{Andrewsetal09,Eisneretal08,AndrewsWilliams07a,Kitamuraetal02}) the
colder regions outside of $\sim 50$ AU are if anything more likely to show
evidence of gravitational instability that at the radii we simulate, and
indeed any spiral structures forming at large radii will be more easily
resolvable.  In this sense therefore our predictions may even be conservative
in terms of the maximum distance at which spiral structures may be detectable.

Likewise our cooling prescription, although simplistic, is valid for regions
at large radii where the temperature is below the ice sublimation
temperature, and is therefore not an unreasonable simplification. It
should be noted that the ratio of the cooling time to the dynamical
time $\beta=t_{\rm cool}\Omega$ determines the amplitude of the spiral
perturbation and hence the contrast in the simulated images. The value
$\beta=7$ adopted in this paper is in the right range for discs at a
few tens of AU. However, a larger value of $\beta$ (that is, a less
efficient cooling) would provide a relatively smaller contrast in the
ALMA images.  

Finally, note that in this paper we have considered the contribution
to the sub-millimetre emission due solely by the disc. In the earliest
phases of star formation, the system might show a substantial emission
on larger scales, produced by the infalling envelope feeding the
disc. This larger scale contribution has been neglected in the present
paper.  

As noted in the Introduction there are already possible detections of spiral
structures in the discs of GSS 39 in Ophiuchus \citep{Andrewsetal09} and in
IRAS 16293-2422B \citep{Rodriguezetal05}.  However the structures in GSS 39
are not robust at the 3$\sigma$ level \citep{Andrewsetal09}, and those in IRAS
16293-2422B, whilst appearing to be genuine, may plausibly be due to
interactions with a companion.  Furthermore, the presence of
planetary mass objects within the disc may also lead to the
excitation of spiral density waves \citep[see for 
instance][]{GoldreichTremaine80,OgilvieLubow02,ArmitageRice05}.
However in this latter case the one-armed spiral wakes are in general
tightly wound 
(dependent on the disc thickness, \citealt{OgilvieLubow02}) and would
therefore be difficult to detect directly.  Furthermore, analysis by
\citet{WolfD'Angelo05} on the observability of embedded gas giant
planets using ALMA suggests that the spirals would only be visible
very close to the planet itself.  Any detections
of large scale spirals in isolated protostellar discs are therefore
likely to be due principally to the action of the gravitational instability.

Confirmed, unambiguous observations of
gravitationally induced spiral structures within protostellar discs would be
valuable for a number of reasons.  As the gravitational instability is
expected to operate during the early phases of star formation, processing the
infalling envelope and allowing rapid accretion onto the protostar, such
detections would validate this mechanism for growing the masses of protostars.
More controversially, it may enable models of brown dwarf/low mass stellar
companion formation through disc fragmentation
\citep{BateBB02,StamatellosHW07,StamatellosW09,Clarke09} to be
validated, as the 
presence (or otherwise) and amplitudes of spiral arms would allow constraints
to be placed on the numbers of such companions that may be expected.  

In a similar vein, detections of spiral features may enable us to determine the
dominant mode of planet formation at various radii, about which there is some
debate.  While the standard core-accretion model is favoured at low radii
\citep{Boley09,Klahr08,Bodenheimeretal00,Lissauer93}, direct
fragmentation of gravitationally induced spirals remains a candidate
mechanism at radii above $\sim 50$ AU
\citep{Boss97,Boleyetal06,Rafikov09,Cossinsetal10}. As the presence of a peak
in the 1.3 cm emission around HL Tau has been put forward as a promising
candidate for a planetary mass companion formed through gravitational
instability, detections of the spiral wave progenitors of such companions
would provide significant backing for this mechanism.  Moreover, the
presence of large amplitude spiral density perturbations may be important for
the formation and growth of planetesimals, both due to the concentration of
the dust fraction within the arms \citep{RiceLPAB04,RiceLPAB06,ClarkeL09}
and further due to the possible scattering of planetesimals by the spiral
potential \citep{BritschCL08}.  In either case, observations of the spiral
arms themselves would place constraints on, and therefore allow us to
discriminate between, the two planet formation modes at large radii.

\section*{Acknowledgements}
\label{acknowledgements}
We acknowledge the use of \textsc{splash} \citep{PriceSplash} for the
rendering of all surface density and optical depth plots.  PJC would
further like to thank Mark Wilkinson for helpful discussions
concerning observational techniques.  
\bibliographystyle{mn2e} 
\bibliography{Cossins}

\end{document}